\documentclass[aip,rsi,amsmath,amssymb,reprint]{revtex4-1}

\usepackage{CJK}
\usepackage{graphicx}
\usepackage{dcolumn}
\usepackage{bm}
\usepackage{fouriernc}		
\usepackage{hyperref}		
  \hypersetup{colorlinks = true, 	
  citecolor = blue,		
  linkcolor = blue, 		
  breaklinks = true,
  backref = true			
  }

\begin{document}
\begin{CJK*}{UTF8}{} 

\preprint{AIP/123-QED}

\title{Construction and evaluation of an ultrahigh-vacuum-compatible sputter deposition source}

\author{Peter Lackner}
\author{Joong-Il Jake Choi}
\altaffiliation[Present address: ]{Center for Nanomaterials and Chemical Reactions, Institute for Basic Science (IBS) and Graduate School of EEWS, Korea Advanced Institute of Science and Technology (KAIST), Daejeon 305-701, South Korea.}
\author{Ulrike Diebold}
\author{Michael Schmid}
\affiliation{Institute of Applied Physics, TU Wien, Vienna, Austria}

\date{\today}

\begin{abstract}
A sputter deposition source for use in ultrahigh vacuum (UHV) is described and some properties of the source are analyzed. The operating principle is based on the design developed by Mayr et al. [Rev. Sci. Instrum. 84, 094103 (2013)], where electrons emitted from a filament ionize argon gas, and the Ar$^+$ ions are accelerated to the target. In contrast to the original design, two grids are used to direct a large fraction of the Ar$^+$ ions to the target, and the source has a housing cooled by liquid nitrogen to reduce contaminations. The source has been used for deposition of zirconium, a material that is difficult to evaporate in standard UHV evaporators. At an Ar pressure of $9\times 10^{-6}$ mbar in the UHV chamber and moderate emission current, a highly reproducible deposition rate of $\approx 1$ monolayer in 250\,s was achieved at the substrate (at a distance of $\approx 50$\,mm from the target). Higher deposition rates are easily possible. X-ray photoelectron spectroscopy shows a high purity of the deposited films. Depending on the grid voltages, the substrate gets mildly sputtered by Ar$^+$ ions; in addition, the substrate is also reached by electrons from the negatively biased sputter target.

\end{abstract}

\maketitle
\end{CJK*}

\section{\label{sec:introduction}Introduction}

Growth of (ultra-)thin films for surface-science experiments in an ultrahigh-vacuum (UHV) environment is usually performed by evaporation, using resistive heating for low to medium temperatures and electron-beam evaporation for materials that require high temperatures for evaporation. For materials that reach a sufficiently high vapor pressure below or at the melting point, electron-beam evaporation from a rod is the method of choice, having the advantage that the material is not in contact with anything that could be a source of contaminations. Other materials require temperatures well above the melting point for sufficient vapor pressures. When the end of a rod gets heated to such a high temperature, a droplet would form and detach from the rod. Thus, such materials (e.g., Al, Au, Sn, Pb) are usually evaporated from crucibles. This requires that the material must not react with the crucible at high temperatures. The crucibles of large electron-beam evaporators (kilowatt range) can be cooled while the melt is at higher temperatures, but this is not the case for small evaporators, where the electrons impinge on the crucible, not the melt. A particularly notorious material for evaporation is zirconium, which has a vapor pressure of only $\approx 4\times 10^{-5}$\,mbar at the melting point ($\approx 1850\,^\circ$C), corresponding to evaporation of about two monolayers (ML) per second; the deposition rate at the substrate is typically lower by four orders of magnitude. Thus, evaporation from rods is very slow. Zr forms eutectic alloys with the refractory metals Mo, Ru, Ta, W, Re, and Ir, which precludes the use of metal crucibles. Zr also reacts with graphite crucibles. Fortunately, these problems are not relevant for sputter deposition.

In many cases, a further advantage of sputter deposition, compared to evaporation, is better layer-by-layer growth of the films. One reason for improved growth is the higher kinetic energy of the sputtered atoms (few eV compared to sub-eV for evaporated atoms) and also the presence of other energetic particles during sputter deposition.\cite{michely_1998} Compared to evaporation, additional advantages of sputter deposition are the possibility to grow compounds with the film composition being close to that of the sputter target, and better stability and reproducibility of the deposition rate compared to most UHV evaporators.

\begin{figure*}
\includegraphics[width=16cm, bb=0 0 473 125]{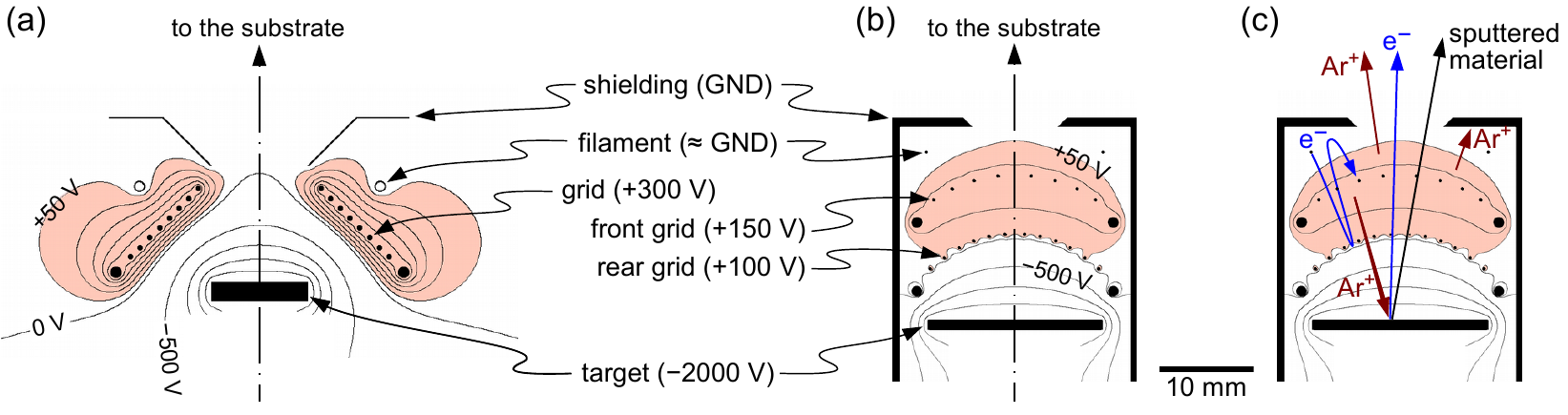}
\caption{\label{fig:potentials} Cross sections and equipotential lines (ignoring space-charge effects) of (a) the original sputter source design by Mayr et al.\cite{mayr_2013,gotsch_2016} and (b) the design presented in this work. Schematic particle trajectories are shown in (c). The light-red shading shows the region where the electron energy exceeds 50\,eV; this is roughly the volume where efficient ionization of Ar is possible. Equipotential lines are drawn in 50\,V intervals for positive and 500\,V intervals for negative voltages.}
\end{figure*}

Conventional magnetron sputter sources using a gas discharge typically operate at pressures between $10^{-3}$ and $10^{-2}$\,mbar and deposition rates of nanometers per second, which is hard to reconcile with UHV surface-science experiments. A way out is using a standard UHV-type ion source for sputtering and collecting the sputtered material at the substrate\cite{michely_1998,schwebel_1986,lee_1996}. This technique is sometimes referred to as ion-beam sputter deposition (IBSD) and requires a special setup of the vacuum chamber. To circumvent the rather low deposition rates achieved with standard (electron-impact ionization) UHV ion sources, high-current ion sources such as duoplasmatron or Kaufman sources can be used.\cite{schwebel_1986,lee_1996} A much simpler approach is a dedicated UHV-compatible sputter deposition source, as recently designed by Mayr et al.\ for deposition of Zr.\cite{mayr_2013,gotsch_2016}.  This source was also found to be useful for oxides, e.g. ZrO$_2$ or YSZ (yttria-stabilized ZrO$_2$). As there is no gas discharge involved, this type of deposition source can be operated at much lower Ar pressures than magnetrons.

The design by Mayr et al.\cite{mayr_2013} comprises a filament at ground potential emitting $\approx 150\,$mA and a grid at $+300$\,V collecting the electrons. When the vacuum chamber is backfilled with Ar, Ar$^+$ ions created by electron-impact ionization are accelerated onto the sputter target biased at $-2$\,keV. At an Ar pressure of $10^{-4}$\,mbar, an ion current of 50--100\,$\mu$A at the target was obtained.\cite{mayr_2013} A closer analysis of the design reveals, however, that the electron cloud ionizing the Ar gas [light red in Fig.\ \ref{fig:potentials}(a)] is mainly at the front side of the grid, where the filament is placed, and at the periphery. The Ar$^+$ ions generated there bombard the shield at the top side of the sputter source and the surrounding vacuum chamber, which may cause deposition of unwanted material onto the substrate and target as well as desorption of molecules into the residual gas. The maximum kinetic energy of the Ar$^+$ ions, when impinging on grounded surfaces, is given by the grid voltage (300\,eV). At the other side of the grid, facing the target, the electrons are repelled by the negative high voltage of the target, so ionization is possible only in a rather small volume close to the grid. Ar$^+$ ions created at this side of grid near the axis are accelerated to the target for sputtering (the desired effect). It should be noted that the number of ions impinging near the target center could be optimized by placing the target somewhat further away from the grid than shown in Fig.\ \ref{fig:potentials}(a); the design in Fig.\ \ref{fig:potentials}(a) is a compromise between optimum sputtering and sufficient heating of the target by thermal radiation from the filament (which is required for sputtering of materials that are insulating at room temperature).\cite{kloetzer_private}

\section{\label{sec:description}Description of the Source}
\subsection{\label{ssec:twogrid}Two-Grid Design}

The problem of only a small fraction of the ionized Ar impinging onto the target can be avoided by using our new design, which has two grids [Fig.\,\ref{fig:potentials}(b); grids are thin dots in the cross section, only the grid holder rings appear as thick dots]. In the region between the two grids, the electric field directs Ar$^+$ ions in the direction towards the target, and the concentric shape of the grids ensures good focusing onto the central area of the target. Nevertheless, some Ar$^+$ ions are also created at the other side of the front grid; these ions bombard either the housing of the source or the substrate, see Fig.\,\ref{fig:potentials}(c). Compared to the original one-grid design,\cite{mayr_2013} the efficiency of the two-grid source is significantly better: With grid voltages of +300 and +200 V for the front and target-side grid, respectively, a target current of 66\,$\mu$A can be achieved with an emission current of $22+12$\,mA (values at the front and rear grid, respectively). This is roughly a quarter of the emission current needed for the same ion current in the one-grid design, at a much lower Ar gas pressure in the UHV system. We used a pressure of $9\times 10^{-6}$\,mbar Ar in the vacuum chamber. 
As the Ar leak valve is connected to the inside of the source housing, the pressure in the source is 27 times higher, about a factor of 2--3 above the pressure used by Mayr et al.\ ($\approx 10^ {-4}$\,mbar in the UHV chamber, equal to the pressure in the source\cite{mayr_2013}). 
The ratio of source and chamber pressures can be determined either from the ratio between the pumping speed and conductance of the orifice or from comparing the target current with Ar supplied to the source and that obtained by backfilling the chamber with Ar from a leak valve not connected to the source.  These two values agree very well ($<3\,$\% difference). 

We can also use lower grid voltages of +150 and 100\,V, as shown in Fig.\ \ref{fig:potentials}(b). This requires a higher filament current (2.4\,A through a 0.15\,mm diameter W wire) and yields a target current of 66\,$\mu$A at 27\,mA emission current (thereof 18\,mA into the front grid and 9\,mA into the back grid).  At these conditions, the total power dissipated by electron impact on the grids is only 3.6\,W, compared to 45\,W in the original one-grid design.  We have also tried higher gas pressures and found that the ion current is proportional to the Ar pressure up to $\approx 2.5\times 10^{-5}$\,mbar (corresponding to $\approx 7\times 10^{-4}$\,mbar in the source); at higher pressures the efficiency (ion current at the target per emitted electron and gas pressure) decreases.  We could not determine the exact reason for this behavior; possibly it is related to the space charge of the ions.

\subsection{\label{ssec:details}Design Details}

\begin{figure}
\includegraphics[width=8cm, bb=0 0 386 519]{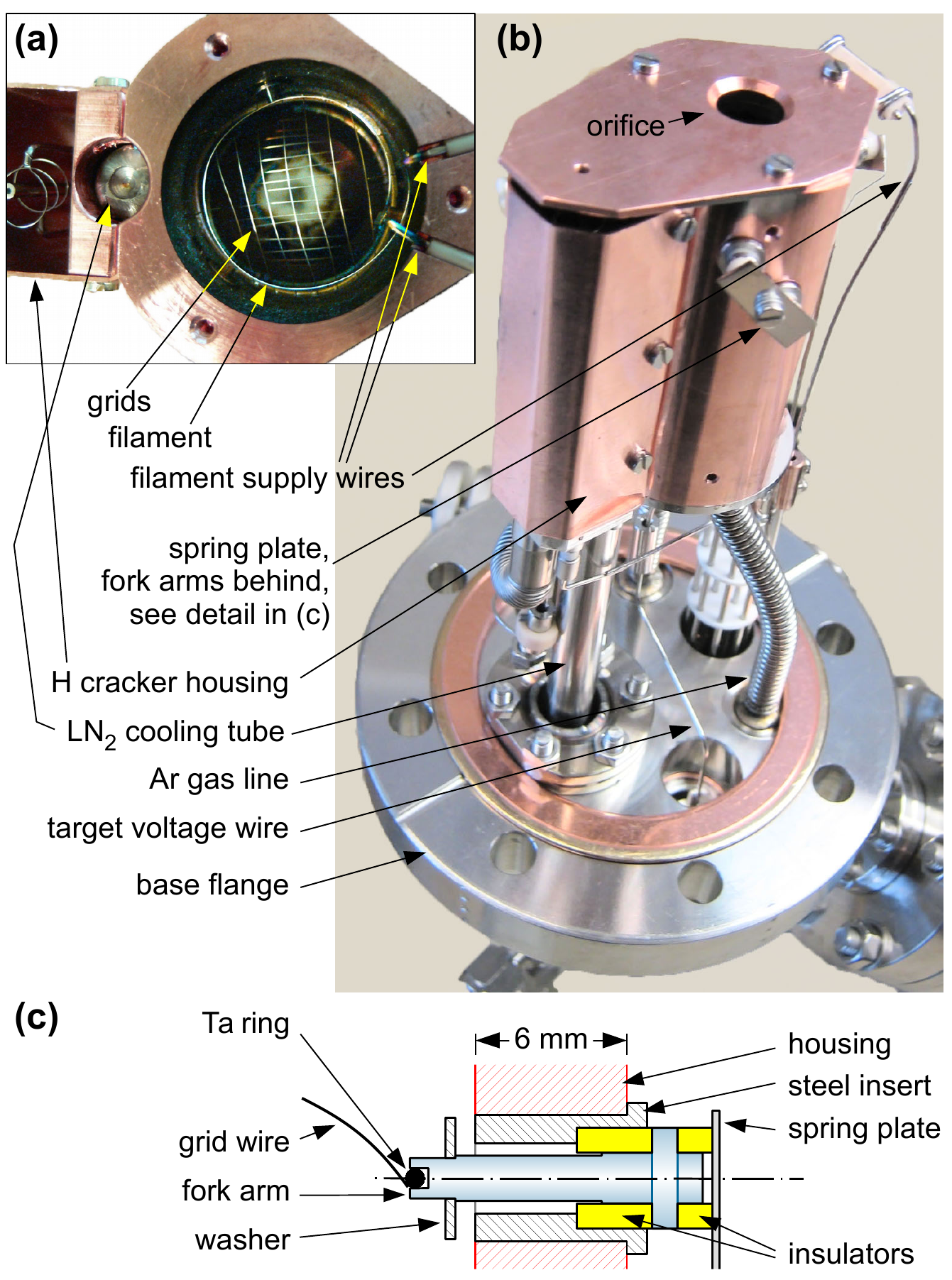}
\caption{\label{fig:photo} (a, b) Photos of the sputter source (the hydrogen cracker also seen is unrelated to the function of the sputter source).  (a) Front view with the top cover removed; note the bright eroded spot in the middle of the otherwise dark target, where the Ar$^+$ ions are focused to (visible behind the grids). (c) Cross section through the housing at the position of one of the insulated fork arms holding a grid.}
\end{figure}

The source region is enclosed by a housing cooled with liquid nitrogen (LN$_2$). 
With deposition of reactive metals (such as Zr) on the inner walls, there is also some pumping of reactive residual-gas molecules (similar to a titanium sublimation pump), which helps to ensure optimum purity of the films deposited. For less stringent purity requirements, the source could also be cooled by water.  
The housing has the shape of a tube with 24\,mm inner and 36\,mm outer diameter, made of a CuCrZr alloy (CW106C) with good thermal conductivity (Fig.\ \ref{fig:photo}). At the outside, this housing has a groove tightly fitting a stainless steel tube with 8\,mm outer diameter (OD) for cooling [Fig.\ \ref{fig:photo}(a)]; a concentric inner tube (3\,mm OD) transports the LN$_2$ to the closed end of the 8\,mm tube. The source housing is pressed against the LN$_2$-cooled tube; this tube also serves as support rod to fix the source on the base flange. The end plate of the source at the front side has a 9\,mm hole (orifice for the sputtered material). The back plate contains a high-voltage feedthrough for the target voltage and the Ar gas inlet, connected by hydraulically formed bellows to the weld in the base flange where the tube from the leak valve enters.

The filament is a loop of 0.15\,mm W wire, spot welded to the supply wires (0.8\,mm Ta)
; the negative terminal of the filament supply is at ground potential. 
The grids have the shape of spherical caps, with the center of curvature roughly in the center of the target, to focus the Ar$^+$ ions to the target center [see the bright erosion spot in the center of the target in Fig.\ \ref{fig:photo}(a)]. This yields a uniform deposition rate on the substrate, without vignetting. The grids are made from 0.175\,mm W wire spot-welded onto rings ($\approx 20$\,mm outer diameter) made from 0.8\,mm Ta wire. A previous attempt to make the rings from stainless steel was unsuccessful; at high power (+300\,V at the front grid), the steel ring got hot enough to soften and deform, finally touching the housing of the source.  The rings of the grids are held by fork-like arms (three per grid); these protrude through the source housing and are held by ceramic insulators [Fig.\ \ref{fig:photo}(c)]. 
Washers protect the insulators from sputtered material scattered by the Ar gas. 
For mounting the grids inside the housing, the grids have to be temporarily held in position from the open end of the source housing, then the forks are inserted radially from the outside. Spring plates (one of them marked in Fig.\ \ref{fig:photo}) prevent the forks from sliding back. For Zr deposition, the target is a disk of Zr metal (diameter 18\,mm, thickness 1\,mm) with 99.9\% purity (HMW Hauner, Germany). The target was spot welded to a support rod at its back side; the rod is held by the high-voltage feedthrough in the back plate of the source housing.

The source is placed on a DN63CF base flange (4.5" outer diameter). The base flange has DN16CF (1.25" outer diameter) ports for the electrical feedthroughs, the LN$_2$ tube and the Ar gas. As the source is rather compact, there is enough space for further components on the same base flange.  We have added a hydrogen cracker similar to Refs.\ \onlinecite{bischler_1993,eibl_1998}; the housing of the H cracker is cooled together with the sputter source. In addition, for future extensions, there is some space for mounting further sources, e.g. tiny crucibles for evaporation, at the sides of the sputter source (to make space, two sides of the source housing are milled flat, as visible in Fig.\ \ref{fig:photo}(a).

\section{\label{sec:evaluation}Evaluation of the Source}

\subsection{Deposition Rate}

We can calculate the deposition rate $F$ from the source by assuming a cosine (Lambertian) angular distribution of the sputtered atoms. Assuming an incident ion current $I_\mathrm{sp}$, a sputter yield $Y$, and $r$ being the distance between the target and the substrate, we obtain%
\begin{equation}
   F(r) = \frac{I_\mathrm{sp} Y}{r^2 \pi |e|} = \frac{I_\mathrm{t} Y}{r^2 \pi |e| (1+\gamma)}
   \label{eq:flux}
\end{equation}%
for emission perpendicular to the target, which is the direction to the substrate (Fig.\ \ref{fig:potentials}). The charge of the ions is $|e|$, and the factor $(1+\gamma)$ in the denominator of the last term accounts for the fact that the measured target current $I_\mathrm{t}$ is higher than that of the incident ions ($I_\mathrm{sp}$) due to ion-induced electron emission, with $\gamma$ being the electron yield upon ion impact.
Equation (\ref{eq:flux}) neglects the effects of resonant neutralization of fast Ar$^+$ ions by collisions with neutral Ar. This is justified when considering the short path of the Ar$^+$ ions (1--2\,cm).
With a target current of 66\,$\mu$A, $r = 50$\,mm, $Y= 1.15$ (for 2\,keV Ar$^+ \to \mathrm{Zr}$, Ref.\ \onlinecite{matsunami_1983}), and $\gamma$ in the range of $0.1$--$0.2$ (consistent with the measured electron current at the substrate, see below), Eq.\ (\ref{eq:flux}) yields a deposited flux $F$ between $5$ and $5.5 \times 10^{12}$ atoms per cm$^2$ and second.

Since the deposition rate determined with a quartz crystal microbalance (QCM) is not very accurate in this case (see below), we have determined the deposition rate by submonolayer growth of Zr on a well-defined single-crystal substrate, Rh(111), and determined the area covered by Zr islands by scanning tunneling microscopy (STM). To determine the deposition rate from the Zr-covered area, we assume that the density of Zr atoms in the islands is equal to that in the basal plane of Zr, $1.11 \times 10^{15}$\,cm$^{-2}$. The analysis of the STM images yields a deposition rate of $4.5\times 10^{12}\,\mathrm{cm}^ {-2}\mathrm{s}^ {-1}$. Considering the uncertainty of the sputter yield and the angular distribution, the agreement must be considered excellent. As the quantities in Eq.\ \ref{eq:flux} are not expected to change with time, this also indicates that the sputter source should have excellent stability of the deposition rate, which perfectly agrees with our experience (provided that the sputter target is sufficiently clean, which is usually the case after a few minutes of operation). 
The deposition rate remains also unchanged when the source is operated at an additional oxygen partial pressure of $10^{-6}$\,mbar for growing ZrO$_2$ films (we monitored the reproducibility of the deposition rate by checking for completion of the 5$^\mathrm{th}$ ZrO$_2$ layer with STM; this should be accurate within a few percent).

\begin{figure}
\includegraphics[width=8.38cm, bb=0 0 351 652]{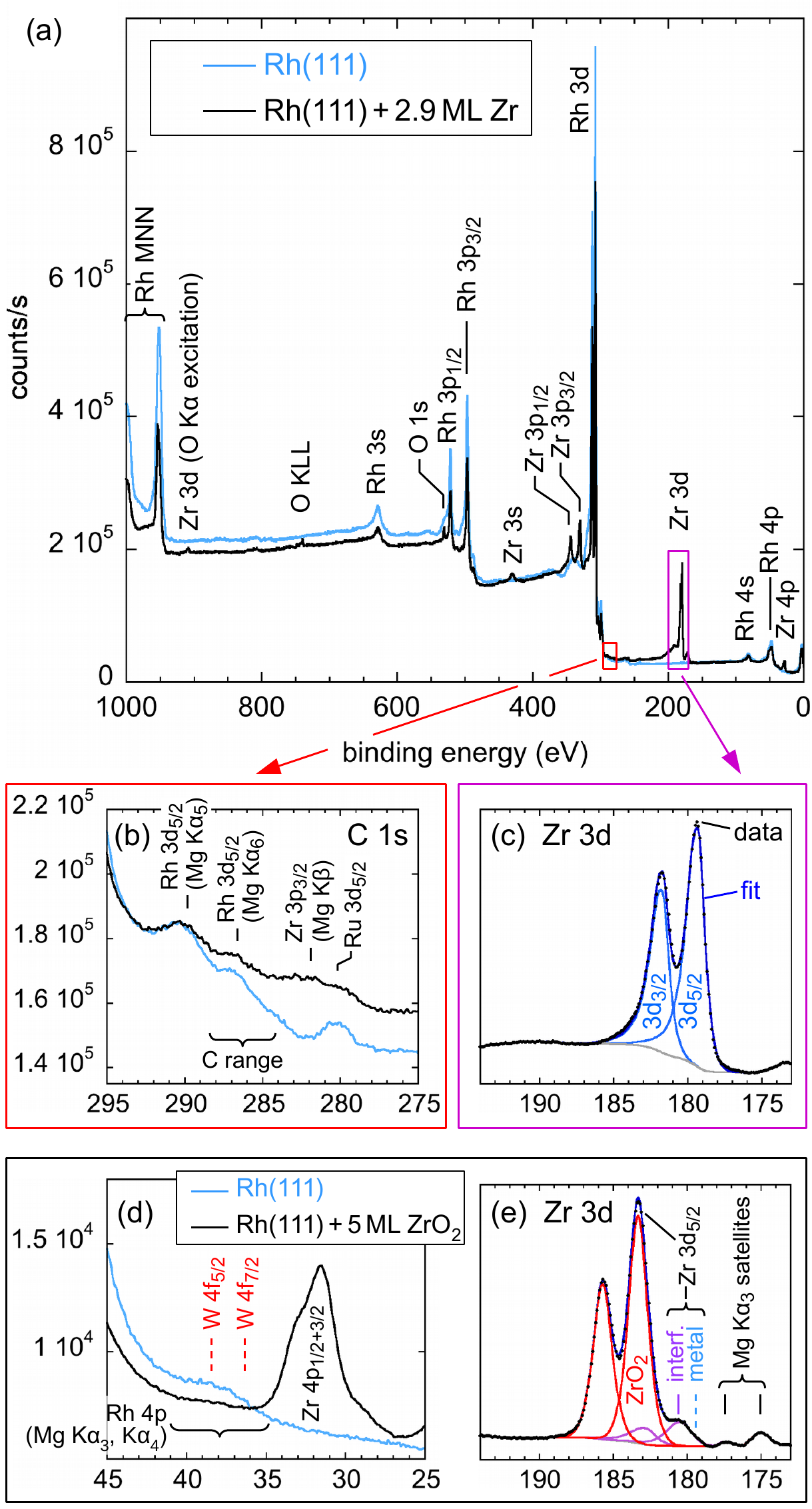}
\caption{\label{fig:xps} (a--c) XPS of a clean Rh(111) surface and after deposition of 2.9\,ML (0.74\,nm) Zr. The C\,1s region (b) was measured with high sensitivity; the range of typical carbon impurities (adventitious carbon and species with C--O bonds) is indicated. The detailed spectrum of the Zr\,3d range (c) shows only metallic Zr (Zr 3d$_{5/2}$ at 179.4\,eV) as indicated by the fit; a slight shoulder to the left, if any, has very low intensity. Spectra obtained after Zr deposition in $10^{-6}$\,mbar O$_2$ [1.0\,nm Zr corresponding to 5\,ML ZrO$_2$(111)] show (d) the absence of (oxidized) tungsten and (e) only oxidic Zr peaks. Analyzer pass energy (a) 20\,eV (b) 50\,eV (for increased count rate); in all other spectra 16\,eV. }
\end{figure}

\subsection{Purity of the Films}

Figure \ref{fig:xps}(a--c) shows x-ray photoelectron spectra (XPS) of a clean Rh(111) surface and immediately after deposition of 2.9\,ML Zr with the UHV sputter source.\cite{argon} We define 1 ML as the areal density of Zr in the basal plane ($1.11\times 10^{15}\,\mathrm{cm}^{-2}$). The Rh(111) substrate was at room temperature during deposition and the measurements. The spectra were acquired with a non-monochromatized Mg K$\alpha$ source and a Specs Phoibos 100 energy analyzer in an analysis chamber coupled to the chamber with the sputter source (the base pressure in both chambers is below $10^{-10}$\,mbar). The only impurity detectable is a small amount of oxygen, probably from oxygen implanted into the Zr target or dissolved there during many previous experiments at $10^{-6}$\,mbar O$_2$ for growing ZrO$_2$ films. The C\,1s region  does not show any indication of carbon-containing impurities; the small peaks found in this range after Zr deposition are due to excitation of stronger lines by satellite lines of the x-ray source, as well as Ru, which is an impurity in our Rh substrate [Fig.\ \ref{fig:xps}(b)]. Zooming in onto the Zr\,3d lines shows only metallic Zr peaks [Fig.\ \ref{fig:xps}(c)]. The Zr\,3d$_{5/2}$ binding energy of 179.4\,eV is slightly higher than usually reported for pure Zr bulk (178.7--178.9\,eV). We attribute this peak shift partly to the surface core level shift, which is about half an eV to higher BE for close-packed surfaces of the early 4d transition metals.\cite{jordan_1990, methfessel_1993} A further contribution will come from interaction of Zr with Rh at the interface, as Zr\,3d$_{5/2}$ energies of $\approx 179.6$\,eV are found for Zr alloys with other late transition metals (Pt$_3$Zr and Pd$_3$Zr, Refs.\ \onlinecite{li_2015,choi_2014}). To some degree, the shift may be also related to interaction with oxygen impurities. In addition, curve fitting cannot exclude an extremely weak shoulder to the left (Zr 3d$_{3/2}$ at 184--184.5\,eV, corresponding to 3d$_{5/2}$ at $\approx 182$\,eV), possibly also related to oxygen: Ultrathin and bulk-like ZrO$_2$ would have the Zr 3d$_{5/2}$ peaks at 180.7 and around 182.8\,eV, respectively.\cite{li_2015}

Figure \ref{fig:xps}(d, e) shows spectra obtained after Zr deposition on Rh(111) at an additional oxygen partial pressure of $10^{-6}$\,mbar at room temperature. At these conditions, a ZrO$_2$ film grows, but without post-annealing the oxide is poorly ordered.  The main peak at 183.3\,eV is from ZrO$_2$ (due to the large band gap of $\approx 5$\,eV, the band alignment and, thus, the binding energies can shift considerably).  There is also a minor peak at 180.4\,eV, which decreases with increasing film thickness, thus it must come from Zr at the interface. As this peak does not disappear upon annealing under oxidizing conditions, we attribute it to ZrO$_2$, not to a lower oxidation state of Zr, in agreement with DFT calculations that predict a lower Zr 3d binding energy at the ZrO$_2$-metal interface.\cite{li_2015}  The spectrum shows no metallic Zr.  Mayr et al.\ have reported that their source, when operated with O$_2$ in the background gas, can lead to tungsten and tantalum impurities in the films, presumably from formation of volatile oxides on the filament and grid materials (W and Ta, respectively; due to the high power these grids get very hot).\cite{mayr_2013} We also checked for tungsten impurities coming from the filament or grid wires (in our source, both are W). Fig.\ \ref{fig:xps}(d) shows no indication of any W signal at the positions where it was observed by Mayr et al.\ (dashed lines in the figure). W would show a sharp doublet there, superimposed on the broad Rh 4p Mg $\mathrm{K}\alpha_3 + \mathrm{K}\alpha_4$ satellite, thus we can exclude such a problem for our source.

\subsection{Ar$^+$ Ion and Electron Emission}

\begin{figure}
\includegraphics[width=8.5cm, bb=0 0 391 204]{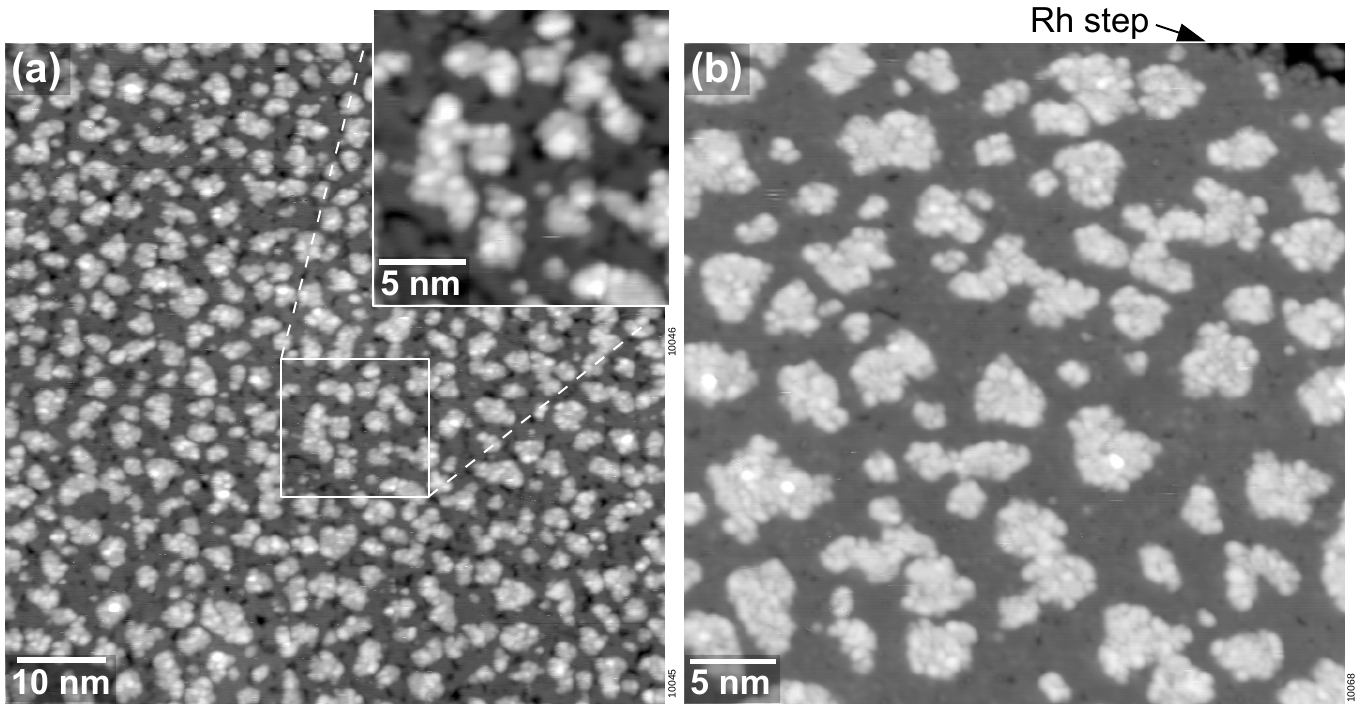}
\caption{\label{fig:stm} STM images of a Rh(111) surface after deposition of 0.35 ML of Zr with front/rear grid voltages of (a) 300 V/200 V and (b) 150 V/100 V. Vacancy islands appear as small black patches in the images. The sputter damage can be reduced drastically by reducing the grid voltages.}
\end{figure}

Apart from sputtered target material, two types of particles are emitted from the source [Fig.\ \ref{fig:potentials}(c)]:  Ar$^ +$ ions and electrons. As mentioned above, Ar gas ionized in the region above the front grid will not be accelerated towards the target but rather to the end plate with the orifice or out of the source, towards the substrate. These ions can sputter material from the inside of the source housing onto the target. Fortunately, after short operation of the source, all the inside of the housing is covered with target material, so there is no contamination of the target by wall material. Some Ar$^+$ ions (with energies in eV up to the voltage of the front grid) also reach the substrate; these Ar$^+$ ions will lead to mild sputtering of the target. Indeed, in our experiments with sputter deposition of Zr onto well-prepared Rh(111) single crystals, we have observed the formation of vacancy islands in the Rh surface.  The scanning tunneling microscopy (STM) images in Figure \ref{fig:stm}(a) show the vacancy islands as small black patches. These vacancy islands can be explained only by sputtering of the sample, which happens in addition to the deposition of sputtered target material. In some cases, this effect can be desired (ion-beam assisted deposition, IBAD)\cite{michely_1998,harper_1985}. Impingement of particles with energies above $\approx 150$\,eV leads to an increased density of nuclei, which can promote layer-by-layer growth in some cases.\cite{schmid_2009} A disadvantage of mild sputtering is reduced accuracy of the deposition rate due to mass removal. When growing oxides or compound materials, preferential sputtering will also alter the composition of the film. For the experiment shown in Fig.\ \ref{fig:stm}(a) the grid voltages were 300 and 200\,V for the front and rear grid, respectively (Ar$^+$ energies up to 300\,eV). We usually choose grid voltages of 150 and 100\,V. In this case, the sputter yield $Y$ at the maximum ion energy (150\,eV) is sufficiently low to avoid these problems (e.g., $Y\approx 0.34$ for 150\,eV Ar$^+$\,$\rightarrow$\,Rh, Ref.\ \onlinecite{matsunami_1983}). This is also seen in Fig. \ref{fig:stm}(b), where only very few and small vacancy islands are found.

The sputter deposition source also emits electrons, which are liberated by ion impact at the target [Fig.\ \ref{fig:potentials}(c)]. These electrons are accelerated to $2$\,keV by the target voltage.
Quantification of the electron current on the substrate is not easy, because only the sum of the electron and Ar$^+$ ion current can be measured; the emission of secondary electrons upon $2$\,keV-electron impact at the substrate further complicates the problem. The substrate current is given by%
\begin{equation}
   I_\mathrm{substr} = I_\mathrm{ion} + I_\mathrm{el}(1 - \delta)
   \label{eq:Isubstrate}
\end{equation}%
where $\delta$ is the secondary electron yield for electron impact at the substrate (due to the low ion energies, ion-induced electron emission from the substrate can be neglected). We have studied this effect with a Pt(111) single crystal serving as substrate. In the initial phase of the deposition, while the substrate was essentially uncovered, we have measured a substrate current of $I_\mathrm{Pt} = 2.2\,\mu$A; with increasing Zr coverage it was found to decrease to $I_\mathrm{Zr} = -1.2\,\mu$A. With these two values, and $\delta = 1.22$ ($0.51$) for Pt (Zr) at 2\,keV,\cite{lin_2005} we obtain an electron current of%
\begin{equation}
   I_\mathrm{el} = -\frac{I_\mathrm{Pt} - I_\mathrm{Zr}}{\delta_\mathrm{Pt} - \delta_\mathrm{Zr}}
   \approx -4.8\,\mu\mathrm{A}
   \label{eq:Iel}
\end{equation}%
and an ion current of $\approx 1.1\,\mu$A. Eq.\ (\ref{eq:Iel}) contains the difference between two secondary electron yield values $\delta$, which are not known with high accuracy; the exact value of $\delta$ may also depend on the crystallographic properties of the material and electron scattering in the layers below (Pt single crystal and Zr thin film in our experiment vs.\ $\delta$ values of polycrystalline material in the literature). Thus, some uncertainty of the electron current is to be expected. This uncertainty has a significant impact on the ion current, which is calculated as the difference of two larger quantities according to Eq. (\ref{eq:Isubstrate}). We have also tried to determine the secondary electron yields with a 2\,keV electron source from the difference between the sample current with and without a positive 30\,V sample bias, yielding $\delta_\mathrm{Pt} = 1.82$ and $\delta_\mathrm{Zr} = 0.68$. These values would result in $I_\mathrm{el}=-3.0\,\mu\mathrm{A}$, but a negative ion current ($-0.2\,\mu\mathrm{A}$). Thus, the ion current cannot be determined with reasonable accuracy  by this method. Based on the sputter damage observed by STM, we estimate that the ion current is actually in between these two values; i.e.\ a few tenths of a $\mu$A. 

Due to the high kinetic energy of the electrons, they dissipate a power of $\approx10$\,mW at the substrate. This is not an issue for most substrates, but it affects the reading of a quartz crystal microbalance used to determine the deposition rate. In our experience, with standard AT-cut 6\,MHz quartz crystals used for thin-film deposition monitors\cite{benes_1995} and a water-cooled crystal holder, the time for equilibration (until a stable deposition rate is displayed) is about 10--20\,min for thermal evaporation, but about an hour with the sputter deposition source (SC-cut quartz crystals, which are also used for sputter yield measurements,\cite{hayderer_1999} are less sensitive to thermal stress and would perform better). In our experience, even after equilibration the QCM readings scatter by more than 10\%. We rather rely on the deposition rate determined once (months ago) from the coverage determined by STM. To ensure best reproducibility, we adjust the gas pressure to get the same target current each time. In other words, the reproducibility of the deposition rate of the sputter source (estimated to be better than a few percent over several months, including several bakeout cycles of the UHV system) is better than that of the QCM readings.

In case that electron or ion emission causes a substantial problem, it would be possible to add an extra electrode at the orifice to repel or deflect these charged species. If the position of the front end of the source is constrained by having to avoid collisions with the sample holder, this would slightly increase the target-to-substrate distance and thereby reduce the deposition rate. If the substrate can be biased, this would provide another possibility to repel either electrons or ions (at the cost of increased kinetic energy of the other species).

\section{Conclusions}

We have presented a sputter deposition source optimized for growth of clean films in ultrahigh vacuum. Compared to the design by Mayr et al.\cite{mayr_2013}, the source operates at more benign values of filament and emission current and lower Ar gas pressures in the UHV chamber for comparable target current and deposition rate. If desired, higher deposition rates could be achieved by increasing the gas pressure or emission current. The source also features excellent long-time reproducibility and the films grown show very high purity. During deposition, the substrate is subject to a flux of both electrons and low-energy Ar$^+$ ions. The energy of the latter can be controlled by the grid voltages, so the source provides the possibility of either ion-beam-assisted deposition or negligible sputtering of the substrate and films by Ar$^+$ ions. Apart from materials that are difficult to evaporate in UHV (such as Zr), due to the high purity of the films grown we consider our source a good choice also for many other materials, especially when considering the advantages of sputter deposition mentioned in the introduction.

\begin{acknowledgments}
The authors would like to thank Peter Varga for helpful discussions, Herbert Schmidt and Rainer G\"artner for the machine shop work and fruitful discussions about construction details, and Martin Leichtfried for designing some of the parts and assembling the source.  This work was supported by the Austrian Science Fund (FWF) under project number F4505 (Functional Oxide Surfaces and Interfaces -- FOXSI).

\end{acknowledgments}

\bibliography{UHV_SputterSource1}

\end{document}